# Towards Maximum Spanning Tree Model in Web 3.0 Design and Development for Students using Discriminant Analysis

S.Padma [1], Dr.Ananthi Seshasaayee [2]

[1] Research Scholar, Bharathiar University, Coimbatore, Assistant Professor, School of Computing Sciences, Vels University, Chennai, India.

[2]Dr.Ananthi Seshasaayee, Associate Professor and Head, Department of Computer Science Quaid-e-Millath Government College for women, Chennai, India.

**Abstract**

Web 3.0 is an evolving extension of the web 2.0 scenario. The perceptions regarding web 3.0 is different from person to person. Web 3.0 Architecture supports ubiquitous connectivity, network computing, open identity, intelligent web, distributed databases and intelligent applications. Some of the technologies which lead to the design and development of web 3.0 applications are Artificial intelligence, Automated reasoning, Cognitive architecture, Semantic web. An attempt is made to capture the requirements of Students inline with web 3.0 so as to bridge the gap between the design and development of web 3.0 applications and requirements among Students. Maximum Spanning Tree modeling of the requirements facilitate the identification of key areas and key attributes in the design and development of software products for Students in Web 3.0 using Discriminant analysis.

## 1. Introduction

Web 3.0 is an extension of www, in which the information can be shared and interpreted by other software agent to find and integrate applications to different domains. Web 3.0 provides integrated real time application environment to the user. The applications are involved in searching using semantic web, 3D web and are media centric. Web 3.0 supports pervasive components. Each component and its relations are represented below.

In web 3.0, web is transformed into database or Data Web wherein the data which are published in the web is reusable and can be queried. This enables a new level of data integration and application interoperability between platforms. It also makes the data openly accessible from anywhere and linkable as web pages do with hyperlinks. Data web phase is to make available structured data using RDF[1]. The scope of both structured and unstructured content would be covered in the full semantic web stage. Attempts will be to make it widely available in RDF and OWL semantic formats.

The driving force for web 3.0 will be artificial intelligence. Web 3.0 will be intelligent systems or will depend on emergence of intelligence in a more organic fashion and how people will cope with it. It will make applications perform logical reasoning operations through using sets of rules expressing logical relationships between concepts and data on the web. With the realization of the semantic web and its concepts web 3.0 will move into Service Oriented Architecture.

The evolution of 3D technology is also being connected to web 3.0 as web 3.0 may be used on massive scale due to its characteristics.

Web 3.0 is media centric where users can locate the searched media in similar graphics and sound of other media formats.

The pervasive nature of web 3.0 makes the users of web in wide range of area be reached not only in computers and cell phones but also through clothing, appliances, and automobiles.

## 2. Review of Literature

Claudio Baccigalupo and Enric Plaza discussed in the paper poolcasting : a social web radio architecture for Group Customization about Pool casting a social web radio architecture in which groups of listeners influence in real time the music played on each channel. Pool casting users contribute to the radio with songs they own, create radio channels and evaluate the proposed music, while an automatic intelligent technique schedules each channel with a group customized sequence of musically associated songs[2] . M.T.Carrasco Benitez discussed in the paper Open architecture for multilingual social networking about an open architecture for all the multilingual aspects of social networking. This architecture should be comprehensive and address well-trodden fields such as localization, and more advanced multilingual techniquesto facilitate the communication among users[3] .

Autona Gerber, Alta van der Merwe, and Andries Barnard discussed in the paper A functional Semantic web architecture about the CFL architecture which depicts a simplification of the original architecture versions proposed by Bernes-Lee as a result of the abstraction of required functionality of language layers. Gerber argues that an abstracted layered architecture for the semantic web with well defined functionalities will assist with the resolution of several of the current semantic web research debates such as the layering of language technologies [4].

Ferda Tartanoglu val'erie Issarny, Alexander Romanovsky and Nicole Levy discussed in the paper Dependability in the web services architecture which discusses about how to build dependable systems based on the web services architecture. It surveys base fault tolerance mechanisms and shows how they are adapted to deal with the specifics of the web in the light of ongoing work in the area[5]. Barry Norton, Sam Chapman and Fabio Ciravegna discussed in the paper developing a Service- Oriented Architecture to Harvest information for the Semantic web which discusses about the Armadillo architecture, how it is reinterpreted as workow templates that compose semantic web services and show how the porting of Armadillo to new domains, and the application of new tools, has been simplified[6].

## 3. Problem Definition

The Design and Development of web 3.0 products are on the course. Due to the existence of the ambiguity in the requirements of Students for structuring the web 3.0 products , bridging the gap between web 3.0 design and development fraternity and Students becomes need of the hour. The key factors for students are to be identified and their preference order is to be extracted.

Let G1, G2, G3 denote the three groups for design and development in web 3.0 . The problem is to find the order of preferences in the three groups and their attributes for Students by identifying the attributes $v_1$ , $v_2$ , ….. $v_n$ included and hence design Maximum Spanning Tree based model.

### 3. Materials and Methods

We collected the perceptions of students inline with web 3.0 attributes. A five point scale was followed which ranges from very low satisfaction , low satisfaction, Medium satisfaction, high satisfaction to very high satisfaction.

1. Block diagram of Web 3.0 Maximum Spanning Tree Model

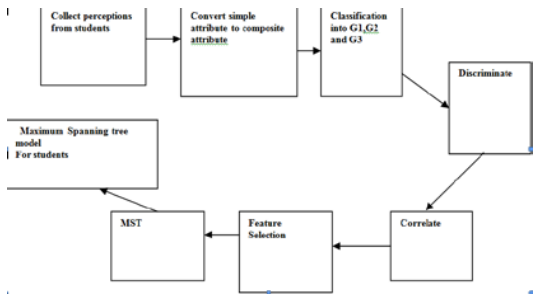

Table 1 : Simple to Composite Attributes in Media

| S.No | Simple Attribute | Composite Attribute |
|---|---|---|
| 1. | 2D Videos, Screen Partial 2D and Partial 3D | 2D |
| 2. | 3D text, 3D screen, 3D demo and 3D tutorial website | 3D |

## 4. Algorithm in Web 3.0 Spanning Tree modeling

a. Start
b. Collect the perceptions regarding the attributes of web 3.0 for students.
c. Combining simple attributes v1,v2…vn into composite attribute p1,p2,….pk.
d. Classification of composite attributes p1,p2,…pk into three groups G1, G2 and G3.
e. Discriminant Modeling for the Students
f. Correlation Coefficient among the composite attributes p1,p2….pn in groups G1, G2
g. Feature selection on composite attributes
h. Maximum Spanning Tree Algorithm.
i. Design maximum spanning tree model for students

**Preprocessing**

The data collected are verified for completeness. The missing values are replaced with the mean value.

**Transforming from Simple attributes to composite attributes**

Data's are collected based on the attributes. Some attributes are composite which are tabulated below.

Table 2 : Simple to Composite Attributes in Output

| S.No | Simple Attribute | Composite Attribute |
|---|---|---|
| 1. | Mash up, Mash up of results, Mash up of tutorial websites, Mash up of Social networking sites | Result as Mash up |

### 5.3 Classification

The data's are collected from the Students based on the attributes 2D, 3D, Audio, Custom mash up, E decisions, Multilingual, Result as Mash up, Semantic Maps, Semantic Wiki, Software Agents, Speech recognition. Based on the functionality, the attributes they are grouped into G1, G2 and G3. G1 comprises of Multilingual, Semantic maps, Edecisions, Semantic wiki and Software agents . G1 is termed as Applications . G2 comprises of 3D, Audio, 2D and Speech recognition. G2 is termed as Media. G3 comprises of Custom Mash up, Result as Mash up . G3 is termed as Output.

### 5.4 Discriminant modeling on groups:

Table 3 : Classification Function Coefficients for Students

| Group | Students |
|---|---|
| Applications | 14.048 |
| Media | 9.374 |
| Output | 8.074 |
| Constant | -46.475 |

The order of preferences for the three groups are given below based on the above Classification Function Coefficients.

Order of Preferences for Students

| STUDENTS | APPLICATIONS | Multilingual (P1) |
| | | Semantic map(P2) |
| | | Semantic Wiki (P3) |
| | | Edecisions (P4) |
| | | Software agents (P5) |
| | MEDIA | 2D (P6) |
| | | 3D (P7) |
| | | Speech recognition (P8) |
| | | Audio (P9) |
| | OUTPUT | Custom mash up (P10) |
| | | Result as mash up (P11) |

From the above table the design and development of web 3.0 products specifically related to Students, can ensue the preference orders and attributes . The products can be designed with the maximum attributes in the first group preference followed by lesser attributes in the second and third group.

### 5.5 Correlation Coefficient between all pairs of composite attributes

The correlation coefficient for all pairs among the Groups are calculated using the following formula.[7]

$$\text{Correlation}(r) = [N\Sigma XY - (\Sigma X)(\Sigma Y)] / \sqrt{[N\Sigma X^2 - (\Sigma X)^2][N\Sigma Y^2 - (\Sigma Y)^2]}$$

where
    $N$ = Number of values or elements
    $X$ = First Score
    $Y$ = Second Score
    $\Sigma XY$ = Sum of the product of first and Second Scores
    $\Sigma X$ = Sum of First Scores
    $\Sigma Y$ = Sum of Second Scores
    $\Sigma X^2$ = Sum of square First Scores
    $\Sigma Y^2$ = Sum of square Second Scores

sample correlation coefficient of G3 (output) for Students

Table 4 : Correlation Coefficient

| S.No | Source | Destination | Correlation Coefficient |
|---|---|---|---|
| 1. | p10 | p11 | .139 |

### 5.6 Feature Selection on Composite Attributes:

The Attribute pairs which have positive correlation are selected. The Attribute pairs which have negative correlation are removed.

### 5.7 Maximum Spanning Tree Algorithm

A spanning tree of an undirected graph of n nodes is a set of n − 1 edges that connects all nodes. This note develops two algorithms for finding the minimum spanning tree. Properties of spanning trees In a spanning tree:

• There is no cycle: a cycle needs n edges.

• There is exactly one path between any two nodes: there is at least onepath between any two nodes because all nodes are connected. Further,there is not more than one path between a pair of nodes because thenthere would be a cycle that includs both nodes.

• Adding a non-tree edge creates a cycle: Suppose a non-tree edge (x, y) is added to a spanning tree. Now there are two distinct paths between(x, y), the added edge and the path in the tree. Hence there is a cycle.

• Removing an edge from a cycle as above creates a spanning tree: after removal of the edge there are (n − 1) edges. All nodes of the graph are connected: suppose edge (x, y) is removed that belonged to the original graph. The nodes x, y are still connected because x, y were on a cycle. For other node pairs, in the path in the original graph replace the edge (x, y) by the path between x, y.

**Kruskal's Algorithm**

Let $G = (V, E)$ be the given graph, with $|V| = n$
{

Start with a graph $T = (V, \phi)$ consisting of only the vertices of $G$ and no edges;

/* This can be viewed as $n$ connected components, each vertex being one connected component */

Arrange E in the order of increasing costs;

for ($i = 1, i <= n - 1, i + +$)

{

Select the next biggest cost edge;

if (the edge connects two different connected components)

add the edge to $T$;

}

}

Order of preferences for Groups and Attributes for students :

| Students | 1. Applications | 1.1 Multilingual (p1) | |
| | | 1.2 Semantic Maps (p2) | 0.297 |
| | | 1.3 E decisions (p4) | 0.273 |
| | | 1.4 Semantic Wiki (p3) | 0.268 |
| | | 1.5 Software agents (p5) | 0.206 |
| | 2. Media | 2.1 3D (p7) | |
| | | 2.2 Audio (p9) | 0.49 |
| | | 2.3 2D (p6) | 0.143735 |
| | | 2.4 Speech recognition (p8) | |
| | 3. Output | 3.1 Custom Mash up (p10) | |
| | | 3.2 Result as Mash up (p11) | 0.139 |
| | | Total cost | 1.816735 |

Maximum Spanning Tree model for Students

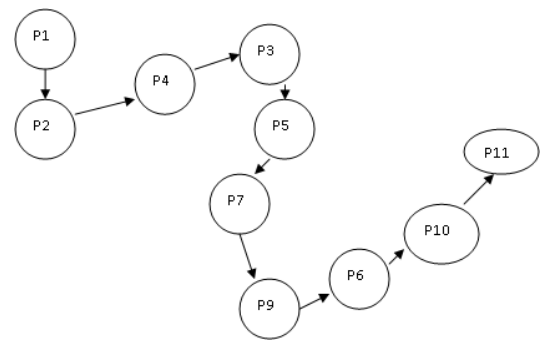

## 7.0 Conclusion

The perceptions inline with web 3.0 products are collected from students. The data's are preprocessed , classified, Mean, Standard deviation and correlation coefficient are computed to understand the descriptive and Discriminant modeled. An model for Students based on Maximum Spanning Tree is designed . At the outset of evolving growth in Web 3.0 this model is an initiative for the of web 3.0 product design for Students .

## AUTHORS PROFILE

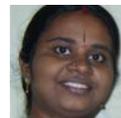

**S.Padma** , is a research scholar in Bharathiar university ,Coimbatore. She has published 2 international journals . Her area of interest is web mining.

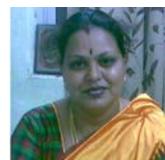

**Dr. Ananthi Seshasaayee** received her Ph.D in Computer Science from Madras University. At present she is working as Associate professor and Head, Department of computer science, Quaid-e-Millath Government College for Women, chennai. She has published 16 international journals. Her area of interest involve the fields of Computer Applications and Educational technology.